\documentclass{article}
\usepackage[utf8]{inputenc}
\usepackage{fancyhdr}
\usepackage{lastpage}
\usepackage{setspace}
\usepackage{lipsum}
\usepackage{geometry}
\usepackage{amsmath,amsthm,amsfonts,mathrsfs}
\usepackage{graphicx}

\usepackage{algorithmic}
\usepackage{multirow}
\usepackage{multicol}
\usepackage{appendix}
\usepackage{indentfirst}
\usepackage{calc}

\usepackage{threeparttable}
\usepackage[capposition=top]{floatrow}
\usepackage{algorithm}

\usepackage{titlesec}

\usepackage{hyperref}

\newgeometry{top = 1.5in, bottom = 1.5in, outer = 0.75in, inner = 0.75in,}
\hypersetup{
    colorlinks=true,
    linkcolor=blue,
    filecolor=magenta,      
    urlcolor=cyan,
    pdftitle={Overleaf Example},
    pdfpagemode=FullScreen,
}

\begin{document}

\begin{flushleft}
{\Large
\textbf\newline{Optimization of COVID-19 Vaccination Process Using Queueing Theory .}
}
\newline
\\
Sarkis Der Wartanian\textsuperscript{1,*},
Baydaa Al Ayoubi\textsuperscript{2},

\bigskip
\bf{1}-Department of Statistics and Applied Probability, University of California Santa Barbara, CA, USA
\\
\bf{2}-Department of Applied Mathematics, Faculty of Sciences, Lebanese University, Hadath, Beirut, Lebanon
\\
\bigskip
* sarkisderwartanian@ucsb.edu

\end{flushleft}

\section*{Abstract}
This research is about COVID-19, which is a contagious virus that reached many countries, including Lebanon. Monitoring the outbreak, researchers have been involved in introducing COVID-19 targeting vaccines. Already facing financial and political issues, Lebanon was further affected by the COVID-19 outbreak. The hereby research tends to design the optimum vaccination plan to reach the recommended herd immunity in Lebanon. The scheme is drafted per application of the queue modeling where the essential parameters revolve around the arrival rate (the mean of which is 536 people) and service rate. Based on a two-stage cluster sampling method, this data is collected from six vaccination centers located in Beirut. The outcomes were computed through R statistical software after validating the stability condition. Per the results, the mean number of people in the network queue ranges between 4.1 and 5.2. In contrast, the time spent by individuals extends from around 5.7 up to 8 minutes, having an average of 5 servers for system one and 4 servers for system two. Therefore, the optimization of the vaccination plan followed in Lebanon necessitates the consistency of arrivals and the decrease of the service time. Furthermore, the service rate can be maximized by adding servers on both systems of the network queue. Further studies might include applying queue modeling in other medical sectors, for instance, the emergency section of hospitals.

\pagestyle{fancy}
\lhead{Article written by Sarkis Der Wartanian}

\newcommand{\Prob}[1]{\mathbb{P}\left(#1\right)}
\newcommand{\E}{\mathbb{E}}
\newcommand{\I}{\mathbb{I}}
\newcommand{\PP}{\mathbb{P}}
\newcommand{\Q}{\mathbb{Q}}
\newcommand{\R}{\mathbb{R}}
\newcommand{\T}{\mathscr{T}}
\newcommand{\SET}[1]{\left\{#1\right\}}
\newcommand{\X}{\mathscr{X}}
\newcommand{\Probs}[2]{\mathbb{P}_{#1}\left(#2\right)}
\newcommand{\EPS}{\varepsilon}
\newcommand{\FI}{\mathscr{I}}
\newcommand{\RANK}[1]{\text{rank}\left(#1\right)}

\newcommand{\BAR}[1]{\overline{#1}}

\newtheorem{theorem}{Theorem}
\newtheorem{lemma}{Lemma}
\newtheorem*{remark}{Remark}

\setcounter{secnumdepth}{3}
\newpage

\tableofcontents

\section{Introduction}

On December 8, 2019, a group of 41 patients from the Huanan South China Seafood Market in Wuhan City documented alarming symptoms \cite{gralinski2020return}. By June 21, 2021, the virus has spread to 222 countries, infecting 179,419,204 individuals in total, with a death toll of 3,885,375 \cite{worldometer}. Nonetheless, the contagion has now been put on the verge of withholding by the evolvement of vaccines.
Accordingly, the optimal prioritization allocation shall be subordinate to the age, comorbidity, susceptibility \& exposure variables, depending on vaccines within each country \cite{bubar2021model}. 

Mass vaccination of the population essential to reach the required herd immunity compels optimum monitoring and management. This paper aims to devise the optimal immunization prototype, improving the already-established vaccination plan and accelerating its course in Lebanon specifically.
For this, the queue modeling is put in service of data analysis, aiming to identify the insipid points of the vaccination program and to highlight the aspects that contribute to better results in optimizing the arrival and service rates of the vaccination systems.

\section{Literature Review}

SARS-CoV-2 having a high transmission rate became a pandemic and resulted in the widespread of the virus worldwide, noting that its fluctuating reproduction number R naught (defined \(\R_0\)) is estimated to range between 1.4 \& 6.49 \cite{liu2020reproductive,safiabadi2021tools}. 

By June 21, 2021, the virus has spread to 222 countries, infecting 179,419,204 individuals in total, with a death toll of 3,885,375 \cite{worldometer}.
However, the contagion has been put on the verge of withholding by the evolvement of vaccines.

An optimal vaccine that pharmaceutical firms attempt to achieve contains singular or multiple antigens, complementary accessories and a dispatching scaffold. Thus, they are altogether targeting the infection in precision, simultaneously being adequate to a broad range of populations and proficient in activating a long-term risk-free immunity \cite{chung2021covid}.

Thanks to the scientific innovations of the century, the clinical trials of the SARS-CoV-2 vaccine began only five months after the first reported infection case \cite{kennedy2020current}.

Considering the characteristics, the SARS-CoV-2 connotes a 67\% herd immunity threshold, suggesting either contamination or host immunization. Unquestionably, infection endangers millions of lives, the reason why all focuses shall shift towards the vaccination of at least 67\% of the global populace, which ensures –whether innate or adaptive immunity response- an immunological memory of the disease \cite{randolph2020herd}.

\subsection{COVID-19 in Lebanon}

Lebanon, a middle-eastern country of only 10,452 km2 with excessively complex demographics, is located in the Mediterranean basin. over an estimate of 6.85 million inhabitants in total by 2019 \cite{worldbank}. Hence, being a Third World Country with a Human Development Index of 0.744 by 2019 \cite{conceiccao2020human}. Lebanon registered its first COVID-19 infection record on February the 21st, 2020 \cite{bizri2021covid}. 
As expected, the first share of vaccines was retrieved in early February 2021, \& Lebanon started administering the vaccines as of February the 10\textsuperscript{th}.
By June 2\textsuperscript{th} 2021, 260,827 individuals have taken different shots, depicting only 5.5\% of the total populace of Lebanon. Nonetheless, it is already conspicuous that the vaccination outcome advocates the expectations, as the daily infections \& death toll have subsided ever since the launching of the vaccination drive, as observed through the below graph \cite{ministryofpublichealth(moph)}.

\begin{figure}[H]
    \centering
    \caption{Daily Numbers of Casualties, Infected \& Vaccinated Individuals}
    \includegraphics[scale = 0.9]{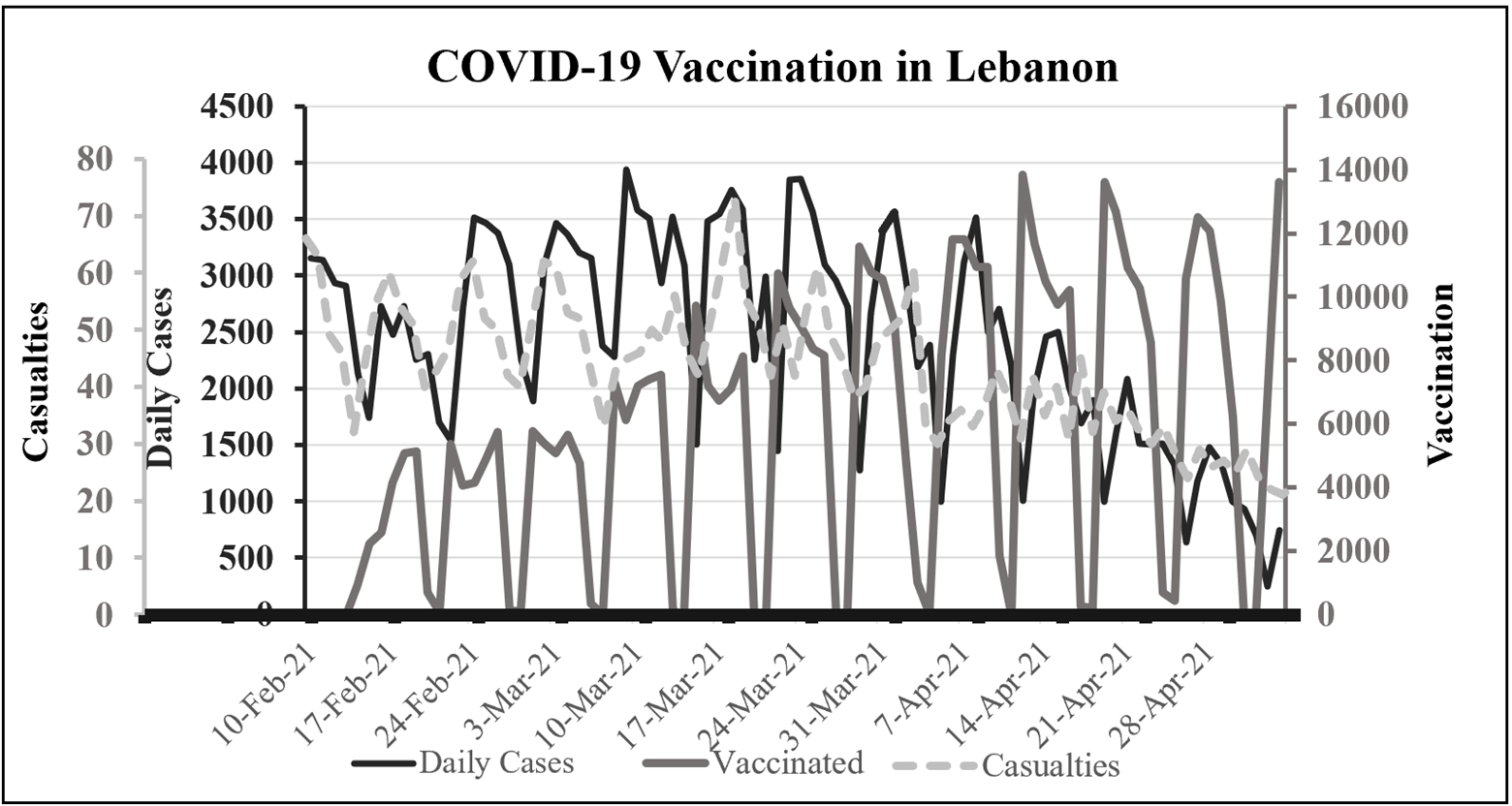}
    \begin{minipage}{10cm}
    \small
    
    Note. Data is collected from February the 10\textsuperscript{th}, 2021 –marking the beginning of the vaccination program- to May 2021.
    \end{minipage}
\label{Figure 1}
\end{figure}

Figure \ref{Figure 1} represents, on a three-axis graph, the daily number of infected individuals in Lebanon, the number of casualties, and the number of vaccinated individuals.
The ground zero would be vaccinating 70 to 80\% of the residents in 2021-2022, ensuring a high leveled herd immunity to sustain the community’s functionality and curtail the burden of the pandemic.

\subsection{Relation Between COVID-19 \& Queue Theory}

According to Mittal, \& Sharma(2020) in \cite{mittal2020deep}, the time spent by patients in the hospital to intake the treatment regarding COVID-19 infection follows a queueing model. More specifically, this model is a multi-server queue model where the inter-arrival times follow a Poisson distribution, and \(\mu \) is the service rate. Moreover, arrivals are served according to first come first served (FCFS) discipline, having more than one server in the system. Lastly, various cases were studied, keeping mu (service rate) and lambda (arrival rate) constant and only changing the number of doctors, who in this event, represent the servers.
Considering the article of Wood and al. \cite{wood2021improving}, the application of queue theory showed to be adequate to pattern the vaccination process in the UK, where M/G/c/\(\infty \)/FIFO model was utilized in the article. Note that M represents the distribution of interarrival-time which is Markov, Exponental; G represents the service-time distribution, where G stands for General; c is the number of servers; \(\infty \) stands for unlimited capacity; and finally the queue discipline is first in first out (FIFO). In this paper, the author considered two scenarios by modifying the arrivals percentage to the vaccination center. As a result, the queue model proves to be adequate in improving the operations of the vaccination centers of the UK.

\section{Methodology}

\subsection{Sampling}

In this article, the author will use the cluster sampling method, which is a probability sampling method. Cluster sampling could be performed in two stages. Throughout the first stage, the population is divided into primary sampling units called clusters. The second stage of cluster sampling was carried out by choosing a smaller unit within each cluster called secondary sampling units. Secondary sampling units are chosen randomly using a simple random sample \cite{innocenti2021optimal}.
In this study, vaccination centers in Lebanon represent the whole population. 
The reason behind choosing Beirut as the first stage cluster and hospitals as the second stage cluster is because around one-third of the populace is vaccinated at Beirut hospitals.
Until June 11, 2021, 339,408 people are vaccinated at Beirut hospitals out of 961,939 as of the total count.
Regarding the population, the inclusion probability is used to determine the primary cluster. This probability gives the percentage increase of including each cluster in the sample.
\begin{equation}
\pi_j = 1- \left( 1 - \frac{N_j}{\sum_{i=1}^nN_j} \right)^k
\end{equation}

where \(\pi_j\) is the probability of cluster j;
\(k \) is the number of chosen clusters.

In this case, \(k \) is 8 (since we have 8 governates.);
\(N_j\) is the  number of individuals contained in cluster j.

\begin{equation*}
\Pi_1 = \frac{339,408}{961,939}=0.353
\end{equation*}

\begin{equation*}
\Pi_2 = \frac{215,914}{961,939}=0.224
\end{equation*}

\(\Pi_1 \) represents the probability of cluster 1, in this case, Beirut. \(\Pi_2\) represents the probability of cluster 2, where cluster 2 stands for Mount of Lebanon.

The highest probability is that of the Beirut cluster, which aggregates more than one-third of the vaccinated individuals' intrapopulation. Mount of Lebanon cluster registers the second-highest probability, 22.4\% of the vaccinated individuals' intrapopulation.
Consequently, the probability random sampling method used is the two-stage cluster sampling, where the 1st stage cluster sample is Beirut, and the 2\textsuperscript{nd} stage cluster sample is hospitals, given that the vaccination centers are all located in the hospitals.

\subsection{Data Collection Method}

Data has been collected from vaccination centers located in the capital of Lebanon, Beirut. Out of eight, six vaccination centers were visited, which are listed below: 
\begin{itemize}
  \item American University of Beirut Medical Center (AUBMC)
  \item Hotel-Dieu de France University Hospital (HDF)
  \item Saint-Georges Hospital University (Roum)
  \item Lebanese American University Medical Center – Rizk Hospital
  \item Lebanese Hospital Geitaoui
  \item Rafic Hariri University Hospital.
\end{itemize}
The collected data is numerical since it constitutes the number of servers existing in each vaccination center and each person's time in the system. System time represents the time spent by a person from the moment of presenting in front of a server before receiving the vaccine, disregarding the observation period after vaccine intake.
Data gathered by the author from the vaccination centers during the June month of 2021. Hence thousands of individuals are observed, through which service time is registered for each individual. 
The gathered data is called primary data since it is supposed during the research period and considered real-time data. Collecting this data is to apply the queueing theory method, identify the input time a person spent in the system, and identify the waiting and service time in the queue.

\subsection{Queueing Theory}

The multi-stage server model is a queueing model used whenever many servers are present in the queue system to serve the arriving people. An example of this queue system is the vaccination centers in Lebanon, where Lebanese are waiting in lines to receive their vaccination dose.
Steady-state is when the arrival rate and the average serving period are the same in the system. Moreover, some measurements' probabilistic manner, including the queue length and people delay in the center, are independent of when the model is studied.
Hence assuming steady-state, the multi-server queueing model perfectly fits the outgoing research, and this model is presented by Kendall’s notation as follows: M/M/c/k/n/discipline \cite{kharel2020steady}. \\
This notation is explained as follows: the first M stands for "Markovian" and indicates that the arrival of customers follows a Poisson process, which means that the time between arrivals is exponentially distributed. The second M also stands for "Markovian" and signifies that the service times are exponentially distributed. This is characteristic of a "memoryless" process, meaning the probability of service completion is independent of how long the service has already been in progress. The number of servers in the system is c, so c servers are working in parallel, each serving one customer at a time. The capacity of the system is represented by k, or the maximum number of customers allowed in the system at any one time, including those being served and those waiting in the queue. If the system is full and a new customer arrives, they will be turned away. Population size is n, it is the total number of potential customers that may require service. If n is infinite, then the customer source is considered unlimited. Finally, discipline refers to the rule used to select which customer is served next. Some examples would be, FCFS (First-Come, First-Served), LCFS (Last-Come, First-Served), SIRO (Service In Random Order), and Priority (customers with higher priority are served first) \cite{harchol2013performance}.

\begin{equation}
\Pi_w={\frac{\left(c\varphi\right)^c}{c!}\left\{\left(1-\varphi\right)\left[\sum_{n=0}^{c-1}\frac{\left(c\varphi\right)^n}{n!}\right]+\frac{\left(c\varphi\right)^c}{c!}\right\}}^{-1}
\end{equation}

\(\Pi_w\) is the probability of the waiting job; delay probability; c is the number of servers in the system; \(\varphi\) is the traffic intensity, and n is the number of population which is finite.

A queueing network is a composition of many systems where the exits from some systems form the arrivals of others. The network is assumed closed if the total amount of people entering the network is fixed, and no external departures are permitted.
Hence, assuming a queue network formed by two systems results in considering the count of arrivals of system two equals the service rate of system one \cite{green2006queueing}.

\begin{figure}[H]
    \centering
    \caption{Network Queue Composed of Two Systems}
    \includegraphics[scale = 0.7]{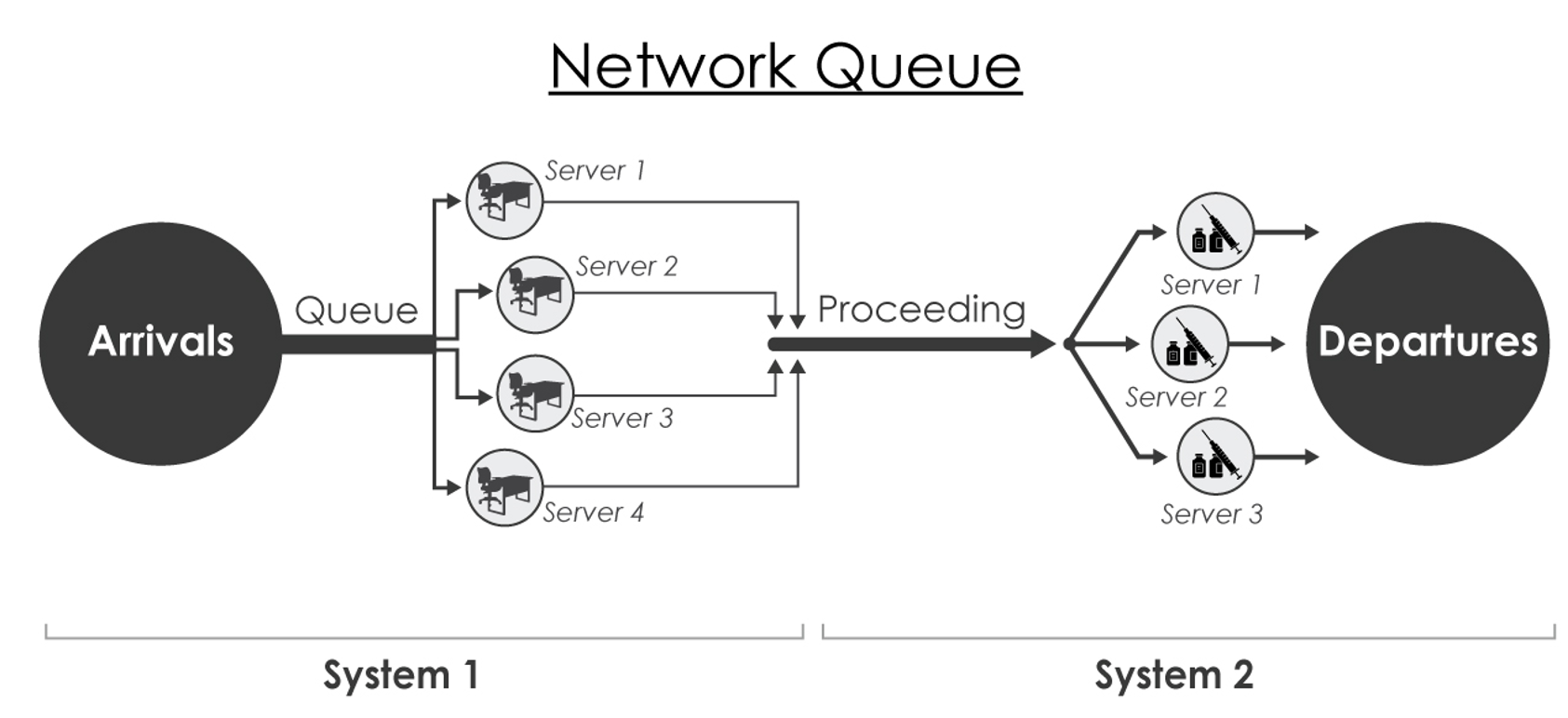}
    \begin{minipage}{10cm}
    \small
    
    Note. This figure is a real example for the queue network of this research, where count of clerks is reduced by one.
    \end{minipage}
\label{Figure 2}
\end{figure}

In figure \ref{Figure 2}, people arriving to system one, are either served by the clerks or waiting in the queue. Finishing from system one, a person proceeds to system two to receive the vaccination dose, however, if all the clerks in system two are busy, this person is delayed, else after receiving the shot, depart from the network.

\subsubsection{Conditions of Queueing Theory} 
To attain statistical equilibrium, the utilization server must fulfill one condition; \(\varphi = \frac{\lambda}{c\mu} < 1\), which ensures that the system is stable, meaning that the queue will not grow indefinitely, that is to say, \(\frac{\lambda}{\mu} < c\), where c is the number of servers as mentioned before, \(\lambda\) is the arrival rate, and \(\mu\) is the service rate. Moreover, supposing to have at least one person or customer present in the model, the system is never idle. Equivalently, the server's operation is never equal to zero (\(\varphi \neq 0\)). \\ 
In general, the traffic intensity \(\varphi\) must not only be less than 1, but it should also be significantly less than 1 to ensure that the system is not perpetually congested. A system with \(\varphi\) very close to 1 may technically be stable, but in practical terms, it would experience long delays. \\
Another condition would be independence between arrivals and services. The next arrival or service completion is not affected by when the previous one occurred.
\cite{adan2015queueing}.

\section{Analysis \& Results} 
\subsection{Sampling Output} 
A person attending the vaccination center has to visit two stations. The first station is where servers check the personal Id of the person undertaking the vaccine and register that person's health information. The second station is where a person intakes the vaccine dose by the servers, mainly nurses.
Consequently, the two stations discussed above are considered as two systems. The queueing model, in this instance, is a recursive model, where the second scheme is a continuation of the initial.

\begin{figure}[H]
    \centering
    \caption{Daily Average Total of People Arriving at the Vaccination Centers}
    \includegraphics[scale = 1.2]{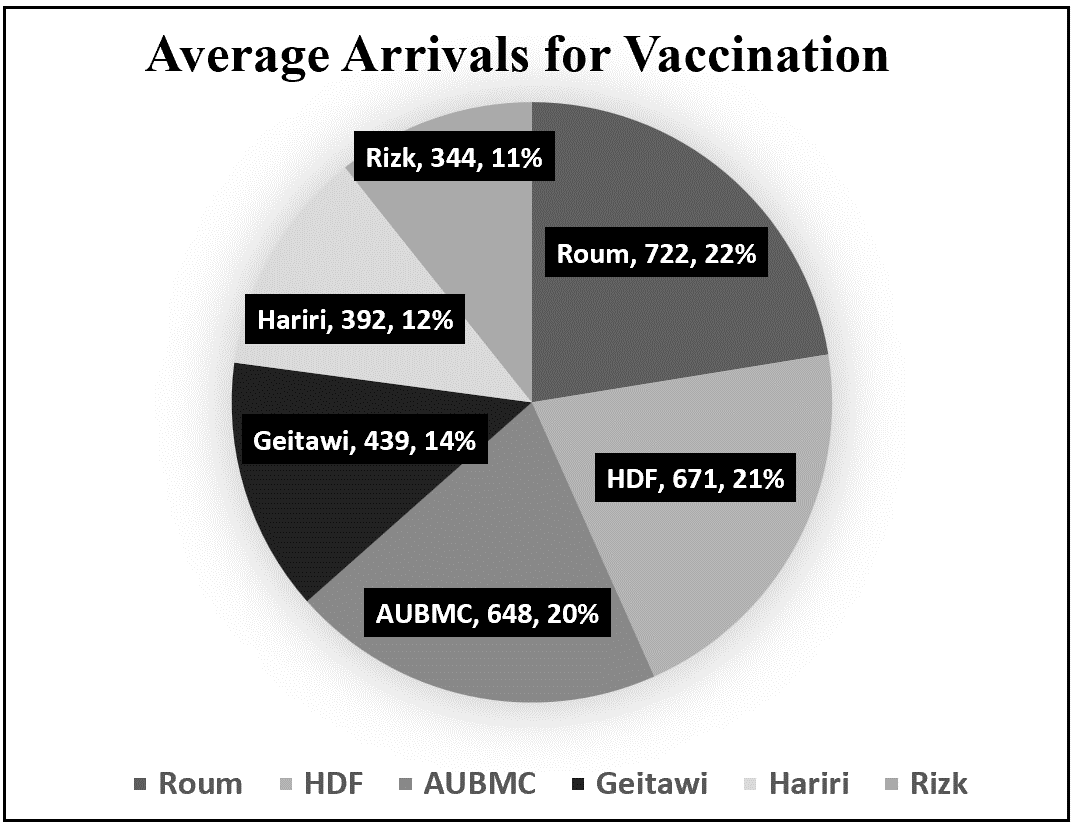}
    \begin{minipage}{10cm}
    \small
    
    Note. Figure 3 represents the average daily arrivals and the percent each center represents from the total.
    \end{minipage}
\label{Figure 3}    
\end{figure}

Figure \ref{Figure 3} shows the six vaccination centers from which data is collected. The most significant arrivals are at “Roum” hospital, achieving 722 mean number of individuals arriving to receive their vaccine. In comparison, the lowest total of appearances is at “Rizk” hospital. Thus, considering the six centers, the mean daily total of appearances is 536 persons.

\begin{figure}[H]
    \centering
    \caption{Mean Count of Servers for Systems One \& Two Inside the Six Vaccination Centers}
    \includegraphics[scale=1.2]{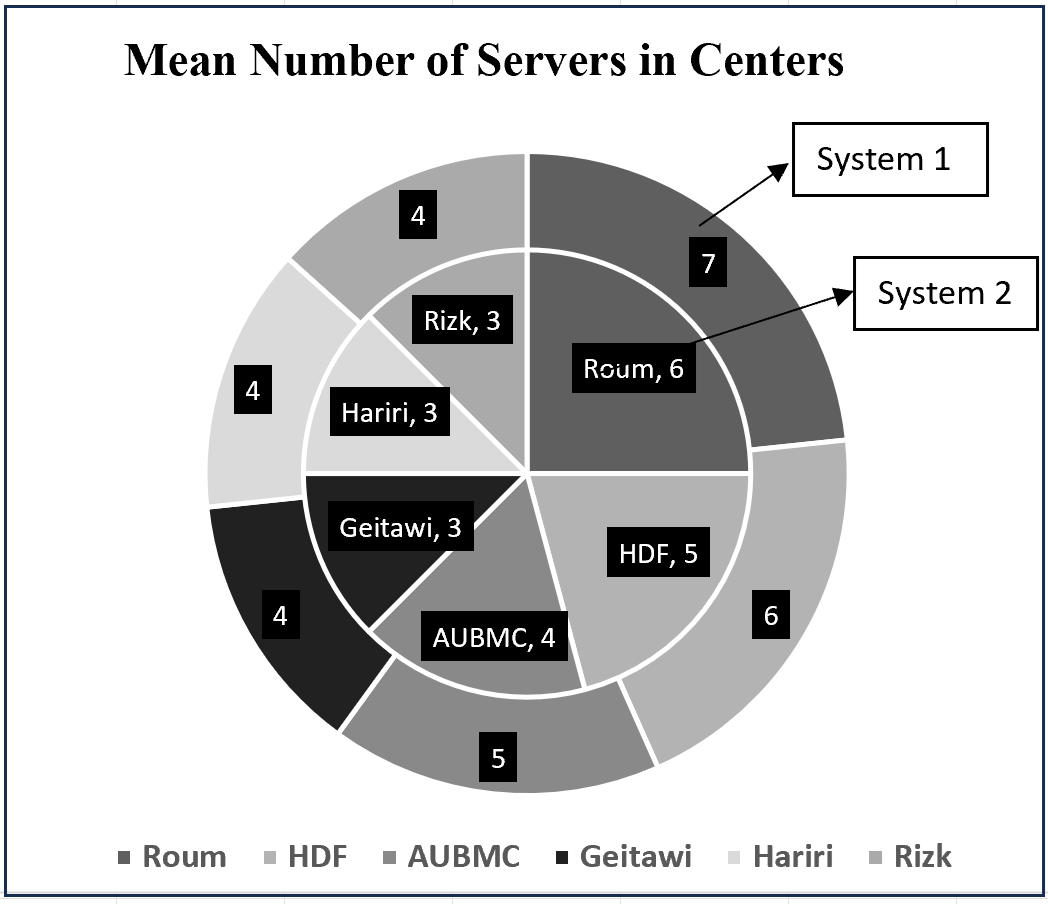}
    
    \begin{minipage}{10cm}
    \small
    Note. This figure shows the mean count of servers for both systems of the queueing network regarding each of the six vaccination centers.
    \end{minipage}
\label{Figure 4}
\end{figure}

Knowing that the queueing model is composed of two systems, figure \ref{Figure 4} shows the count of servers for each system, where the outer part of the pie represents the number of servers of system 1, and the inner part represents that of system 2. Saint George Hospital University Medical Center or “Roum” has the highest percent of arrivals and highest count of servers for both systems.

\subsection{Numerical Computations} 
\subsubsection{Results of System One} 
Data analysis for system one is done using R. The following table summarizes the values of the queueing model variables of system 1:

\begin{table}[H]
\centering
\caption{\bf Computations of System One }
\vspace{1em}
\setlength{\tabcolsep}{25pt} 

\begin{tabular}{|c|c|}

\hline 
Arrival rate = \(\lambda_1\)& 1.117 \\[0.5ex]
Service rate = \(\mu_1\)& 0.409 \\[0.5ex]
Traffic Intensity = \(\varphi_1\)& 0.545 \\[0.5ex]
Count of servers = c& 5\\[0.5ex]
Probability of the awaiting job or cErlang = \(\Pi_w\)& 0.174 \\[0.5ex]
Probability of having nobody in the model = \(P_0\)& 0.063 \\[0.5ex]
Mean queue length = \(\bar{Q}\)&	0.208 \\  [0.5ex]
Long-run number of people in the line \({\bar{Q}}_q\)&	2.199 \\[0.5ex]
Long-run average period spent in queue \({\bar{W}}_q\)&	1.074 \\[0.5ex]
Mean waiting time = \(\bar{W}\) & 	0.187 \\[0.5ex]
Mean service time = \(\bar{S}\)&	2.272 \\[0.5ex]
Mean number of busy servers = \(\bar{n}\)&	2.728 \\[0.5ex]
Mean number of people in the system = \(\bar{N}\)&	2.936 \\[0.5ex]
Mean time spent in the model = \(\bar{T}\)& 	2.629 \\[0.5ex]
Utilization of a single server = \(U_s\)&	0.545 \\[0.5ex]
Utilization of the model = \(U_r\)&	0.937 \\[0.5ex]
Overall utilization = \(U_n\)&	2.728 \\[0.5ex]
     \hline
\end{tabular}
\vspace{1em} 

\begin{minipage}{10cm} 
\small
Note. Source: Author 2021; this table was extracted using Python. It shows the values of the queueing model for system one.
\end{minipage}
\label{Table 1}
\end{table}

Referring to table \ref{Table 1}, it is obvious that system 1 satisfies the stability condition since \(\varphi_1< 1\). Thus, for this multi-server queueing model, the probability of an idle system is 0.063, which means it is unlikely to have no person in the scheme. The mean time of serving a person is around 2.2 minutes, whereas a person's total duration in system one is 2.63 minutes. Besides, the average count of people in system one is three.
Additionally, in the long run, the mean number of individuals in the line is 2.2, whereas the duration of stay in the line is 1.07 minutes. The traffic intensity equals 0.545, which means 54.5\% of the server is used. This value is considered highly acceptable since the higher the utilization level, the longer the delay period in the model.
The table below presents the confidence intervals of the queueing variables computed:

\begin{table}[H]
\centering
\caption{\bf Confidence Interval for Queue Variables of System One}
\vspace{1em}
\setlength{\tabcolsep}{20pt} 

\begin{tabular}{|c|c|c|c|}
\hline 
\multicolumn{2}{|c|}{} & \multicolumn{2}{c|}{\textbf{Confidence Interval}} \\[0.5ex]
\hline
\rule{0pt}{3ex} \textbf{Variables} & \textbf{Values} & \textbf{ Lower Bound }& \textbf{Upper Bound} \\[0.5ex]
\hline
\rule{0pt}{3ex} \(\bar{N}\) & 2.936 & 2.466 & 3.406 \\ [0.5ex]
\(V_N\)& 3.867& & \\[0.5ex]
\(\bar{Q}\)& 0.208& 0.0126& 0.4034 \\[0.5ex]
\(V_{N_q}\)& 0.666 & & \\[0.5ex]
\(\bar{T}\)& 2.629& 2.027& 3.231 \\[00.5ex]
\(V_T\)& 6.331 & & \\[0.5ex]
\(\bar{W}\)& 0.187& 0.042& 0.332\\[0.5ex]
\(V_{T_q}\)& 0.367& &\\[0.5ex]
\hline
\end{tabular}

\vspace{1em} 
\begin{minipage}{12cm} 
\small
Note. Source: Author 2021; this table was extracted using Python. It summarizes the results of the main queueing model variables for the queueing network.
\end{minipage}

\label{Table 2}
\end{table}

As presented in table \ref{Table 2}, the confidence interval of the crucial queueing variables is calculated using the variance of each variable and taking n = 67 since the mean arrival rate is 536 people per day and a day represents 8 hours of work. Also, \(\alpha\) = 0.05 and consequently \(Z_\alpha\) = 1.96 is considered to calculate the critical queue variables in system 2. Knowing the mean count of individuals in the system is 2.936, and using an error margin of 0.47, the confidence interval of \(\bar{N}\) is deduced.
The mean waiting period in the system one line is 0.187 minutes, with a confidence interval of [0.042; 0.332], whereas a person's total period of stay in the scheme is 2.629 minutes, with an error tolerance of 0.6.
After outlining the outcomes of system one, the next part will describe the results of system two of this research.

\subsubsection{Results of System Two}

A person who enters the queueing design and finishes being served in system one automatically continues to plan two to receive the vaccination dose. Data analysis for system two is done using Python, and results are given in the table underneath:

\begin{table}[H]
\centering
\caption{\bf Computations of System Two }
\vspace{1em}
\setlength{\tabcolsep}{25pt} 

\begin{tabular}{|c|c|}

\hline 
Arrival rate = \(\lambda_2\)& 0.409 \\[0.5ex]
Service rate = \(\mu_2\)& 0.244 \\[0.5ex]
Traffic Intensity = \(\varphi_2\)& 0.419 \\[0.5ex]
Count of servers = c& 4\\[0.5ex]
Probability of the awaiting job or cErlang = \(\Pi_w\)& 0.104 \\[0.5ex]
Probability of having nobody in the model = \(P_0\)& 0.184 \\[0.5ex]
Mean queue length = \(\bar{Q}\)&	0.075 \\  [0.5ex]
Long-run number of people in the line \({\bar{Q}}_q\)&	1.72 \\[0.5ex]
Long-run average period spent in queue \({\bar{W}}_q\)&	1.76 \\[0.5ex]
Mean waiting time = \(\bar{W}\) & 	0.1833 \\[0.5ex]
Mean service time = \(\bar{S}\)&	2.324 \\[0.5ex]
Mean number of busy servers = \(\bar{n}\)&	1.676 \\[0.5ex]
Mean number of people in the system = \(\bar{N}\)&	1.75 \\[0.5ex]
Mean time spent in the model = \(\bar{T}\)& 	4.275 \\[0.5ex]
Utilization of a single server = \(U_s\)&	0.419 \\[0.5ex]
Utilization of the model = \(U_r\)&	0.816 \\[0.5ex]
Overall utilization = \(U_n\)&	1.676 \\[0.5ex]
     \hline
\end{tabular}
\vspace{1em} 

\begin{minipage}{10cm} 
\small
Note. Source: Author 2021; this table was extracted using Python. It shows the values of the variables of model two.
\end{minipage}
\label{Table 3}
\end{table}

In this table, the stability condition for system two is satisfied, where \(\varphi_2 < 1\), that is to say \(\frac{\lambda_2}{\mu_2} = 1.67 < c = 4\). Using the probability of waiting job, which is 0.1, various queueing variables are calculated, among which the mean queue length \(\bar{Q}\) = 0.075. In system 2, the probability of having zero people is 0.184, which is almost triple that in system 1. In addition, referring to \(\bar{n}\), out of 4 servers, 1.7 are busy.
Considering the long-run, the average count of individuals in the line is 1.72, whereas the average period spent in the line is 1.76 minutes. Finally, the process of a single server is around 42\%, which is considered highly adequate for being far from the total utilization of value 100\%.
The forthcoming figure sums up the confidence intervals of the queueing variables of system two:

\begin{table}[H]
\centering
\caption{\bf Confidence Interval for Queue Variables of System Two}
\vspace{1em}
\setlength{\tabcolsep}{20pt} 

\begin{tabular}{|c|c|c|c|}
\hline 
\multicolumn{2}{|c|}{} & \multicolumn{2}{c|}{\textbf{Confidence Interval}} \\[0.5ex]
\hline
\rule{0pt}{3ex} \textbf{Variables} & \textbf{Values} & \textbf{ Lower Bound }& \textbf{Upper Bound} \\[0.5ex]
\hline
\rule{0pt}{3ex} \(\bar{N}\) & 1.75 & 1.409 & 2.091 \\ [0.5ex]
\(V_N\)& 2.027& & \\[0.5ex]
\(\bar{Q}\)& 0.075& -0.025& 0.175 \\[0.5ex]
\(V_{N_q}\)& 0.177 & & \\[0.5ex]
\(\bar{T}\)& 4.275& 3.275& 5.275 \\[00.5ex]
\(V_T\)& 17.356 & & \\[0.5ex]
\(\bar{W}\)& 0.183& -0.004& 0.37\\[0.5ex]
\(V_{T_q}\)& 0.611& &\\[0.5ex]
\hline
\end{tabular}

\vspace{1em} 
\begin{minipage}{12cm} 
\small
Note. Source: Author 2021; this table was extracted using Python. It summarizes the outcomes of the main queueing model variables for system two.
\end{minipage}

\label{Table 4}
\end{table}

Referring to table \ref{Table 4}, the mean count of individuals in the line of system two is 0.075, which belongs to the interval [-0.025; 0.175]. The mean waiting time in the queue is 0.183 minutes, which fits in the interval [-0.004; 0.37]. Furthermore, the mean count of individuals in the scheme is 1.75, which belongs to the interval [1.409; 2.091], and the mean duration spent in the model is 4.275 minutes, which belongs to the interval [3.275; 5.275].
The part below represents the network queue composed of the two systems described above.

\subsection{Network Queue}

In this research, the queueing network is composed of two systems. This network is considered a closed queue network since the total count of individuals in the network is constant, and thus no external arrivals are made to the network. Hence, combining the computations of both systems gives a clear image of the variables of the queueing network. 
The table below outlines the results of the following network queue:

\begin{table}[H]
\centering
\caption{\bf Results of the Queueing Network }
\vspace{1em}
\setlength{\tabcolsep}{20pt} 

\begin{tabular}{|c|c|c|c|}
\hline 
\multicolumn{2}{|c|}{} & \multicolumn{2}{c|}{\textbf{Confidence Interval}} \\[0.5ex]
\hline
\rule{0pt}{3ex} \textbf{Variables} & \textbf{Values} & \textbf{ Lower Bound }& \textbf{Upper Bound} \\[0.5ex]
\hline
\rule{0pt}{3ex} \(\bar{N}\) & 4.686 & 4.105 & 5.267 \\ [0.5ex]
\(\sigma_N\)& 2.427& & \\[0.5ex]
\(\bar{Q}\)& 0.283& 0.063& 0.503 \\[0.5ex]
\(\sigma_{N_q}\)& 0.918 & & \\[0.5ex]
\(\bar{T}\)& 6.904& 5.739& 8.069 \\[00.5ex]
\(\sigma_T\)& 4.866 & & \\[0.5ex]
\(\bar{W}\)& 0.37& 0.133& 0.607\\[0.5ex]
\(\sigma_{T_q}\)& 0.989& &\\[0.5ex]
\hline
\end{tabular}

\vspace{1em} 
\begin{minipage}{12cm} 
\small
Note. Source: Author 2021; this table was extracted using Python. It summarizes the outcomes of the main queueing model variables for system two.
\end{minipage}

\label{Table 5}
\end{table}

In this table, the mean number of individuals in the queue is between 0 and 0.5, and the mean number of individuals in the queueing network is 4.7. In contrast, a person's mean waiting time is 0.37 minutes, which belongs to the confidence interval [0.13; 0.6]. Concurrently, the mean duration of time spent in the network varies between 5.74 and 8 minutes.
The probability distributions regarding systems one and two of the original research case considered above are computed in the upcoming section. Moreover, various cases are studied, and the values obtained are compared to those in the research case.

\subsubsection{Ideal Case} 
The values of service and arrival rate are kept similar as in the original case, and only the count of servers is replaced. The one most effective is shown below:
Best Case: \(c_1\) = 7 \& \(c_2 = 6\)
Table 6 represents the values of the parameters of the best case:

\begin{table}[H]
\centering
\caption{\bf Results of Best Case}
\vspace{1em}
\setlength{\tabcolsep}{10pt} 

\begin{tabular}{|c|c|c|c|c|c|}
\hline 
\multicolumn{4}{|c|}{\textbf{Values}} & \multicolumn{2}{c|}{\textbf{Confidence Interval}} \\[0.5ex]
\hline
\rule{0pt}{3ex} \textbf{Variables} & \textbf{System 1} & \textbf{ System 2}& \textbf{Network}& \textbf{Lower Bound}& \textbf{Upper Bound} \\[0.5ex]
\hline
\rule{0pt}{3ex} \(\bar{N}\) & 2.742 & 1.678 & 4.42 & 3.911 & 4.929 \\ [0.5ex]
\(V_N\)& 2.826& 1.694 &4.52 & & \\[0.5ex]
\(\bar{Q}\)& 0.015& 0.003& 0.018& -0.0295& 0.0655 \\[0.5ex]
\(V_{N_q}\)& 0.034 & 0.0054 &0.0394 & & \\[0.5ex]
\(\bar{T}\)& 2.455& 4.099& 6.554 & 5.412 & 7.696 \\[00.5ex]
\(V_T\)& 5.98 &16.759 &22.739 & & \\[0.5ex]
\(\bar{W}\)& 0.013& 0.0075& 0.0205& -0.0204& 0.0614\\[0.5ex]
\(V_{T_q}\)& 0.0153& 0.014 & 0.0293 & &\\[0.5ex]
\hline
\end{tabular}

\vspace{1em} 
\begin{minipage}{12cm} 
\small
Note. Source: Author 2021; this table was extracted using Python. It summarizes the results of the main queueing model variables for the best case.
\end{minipage}

\label{Table 6}
\end{table}

\subsection{Probability Distribution for Queueing Model}

\subsubsection{Original Research Case}

The possibility of waiting more than 1 minute in a queue of system one is: 
\begin{equation}
    \Prob{W>t=1} = 0.068
\end{equation}
Whereas the possibility of spending more than 1 minute in system one is:
\begin{equation}
    \Prob{T>t=1} = 0.72
\end{equation}
The likelihood of waiting more than 2 minutes in a queue of system one is: 
\begin{equation}
    \Prob{W>t=2} = 0.027
\end{equation}
Whereas the likelihood of waiting more than 2 minutes in system one is:
\begin{equation}
    \Prob{T>t=2} = 0.48
\end{equation}
Hence, it is unlikely for a person entering system one to wait more than 1 minute in the queue. However, there is a 72\% chance to spent more than 1 minute in the scheme. This probability is reduced to 48\% for waiting more than 2 minutes in the plan. \newline
Considering system two, the likelihood of waiting more than 1 minute in a queue is: 
\begin{equation}
    \Prob{W>t=1} = 0.059
\end{equation}

Whereas the possibility of spending more than 1 minute in system two is: 
\begin{equation}
    \Prob{T>t=1} = 0.8
\end{equation}
Similarly, the possibility of waiting more than 2 minutes in the queue is computed:
\begin{equation}
    \Prob{W>t=2} = 0.033
\end{equation}
As for the probability of spending more than 2 minutes in the system gives: 
\begin{equation}
    \Prob{T>t=2} = 0.637
\end{equation}
Thus, it is unlikely for a person continuing from system one to system two to wait more than 1 minute, but there is a 63.68\% probability of spending more than 2 minutes in the system. \newline
Consider the Best Case
The table below sums up the outcomes of the best case:

\begin{table}[H]
\centering
\caption{\bf Probability Distribution Regarding the Best Case}
\vspace{1em}
\setlength{\tabcolsep}{20pt} 

\begin{tabular}{|c|c|c|}
\hline 
\multicolumn{1}{|c|}{} & \multicolumn{2}{c|}{\textbf{Best Case}} \\[0.5ex]
\hline
\rule{0pt}{3ex} \textbf{Probabilities} & \textbf{System 1} & \textbf{System 2} \\[0.5ex]
\hline
\rule{0pt}{3ex} \(P\left(W>1\right)\) & 0.004 & 0.0027 \\ [0.5ex]
\(P\left(T>1\right)\)& 0.668& 0.784 \\[0.5ex]
\(P\left(W>2\right)\)& 0.0007& 0.0009 \\[0.5ex]
\(P\left(T>2\right)\)& 0.444 & 0.614 \\[0.5ex]
\(P\left(W>4\right)\)& 0.00002& 0 \\[00.5ex]
\(P\left(T>4\right)\)& 0.196 & 0.377 \\[0.5ex]
\hline
\end{tabular}

\vspace{1em} 
\begin{minipage}{12cm} 
\small
Note. Source: Author 2021; this table was extracted using Python. It summarizes the results of the probability distribution of waiting and system time.
\end{minipage}

\label{Table 7}
\end{table}

In the best case, increasing the count of servers for both systems decreased the probability of waiting for more than several minutes in the queue. Furthermore, the likelihood of spending more than 4 minutes in the system is low and acceptable.

\section{Conclusion} 
This study has effectively demonstrated the use of queueing theory as a vital instrument in optimizing the COVID-19 vaccination process, especially in the challenging context of Lebanon. By meticulously analyzing arrival and service rates at vaccination centers, we have identified key areas for efficiency enhancement, leading to reduced waiting times and a faster overall vaccination process. These improvements, achieved through a methodical, data-driven approach utilizing R and Python, not only streamline the vaccination workflow but also potentially increase public participation in such critical health initiatives. The integration of additional servers in both systems of the queueing network emerges as a pivotal factor, contributing significantly to the increased throughput of the vaccination process.
In managing the complexities of the pandemic, our research underscores the importance of a strategic approach in healthcare logistics, where queueing theory plays a central role. The insights gained here extend beyond pandemic management, suggesting a broader applicability in enhancing patient flow and logistical efficiency in various healthcare settings. As we continue to navigate the ongoing impacts of COVID-19, the adoption of such innovative methodologies in healthcare management becomes crucial, promising more resilient and effective healthcare systems. The addition of specialized vaccination centers, particularly in urban areas like Beirut, would likely expedite the vaccination process further, ensuring a more consistent flow and reducing delays, thereby accelerating the journey towards achieving the necessary herd immunity.

\section{Recommendation} 
In addressing the COVID-19 pandemic, achieving herd immunity through rapid mass vaccination is critical. Key to expediting this process are the optimization of the queueing network, focusing on reducing waiting times and increasing service efficiency, and the expansion of vaccination centers. The efficacy of vaccination relies heavily on maintaining a consistent arrival rate and enhancing the service rate, primarily by increasing the number of servers in each system. For example, adding two servers to each system has shown to significantly improve throughput. Specialized vaccination centers, particularly in proximity to densely populated areas like Beirut, can streamline the process by reducing travel times and stabilizing arrival rates. Innovative solutions like drive-through vaccination and organizing mass vaccination events on non-operational days can further accelerate the immunization drive. Additionally, targeted awareness campaigns in marginalized areas are crucial to improve vaccine acceptance. Training for server clerks to enhance service efficiency is also recommended, as it directly contributes to a reduced service time, thereby accelerating the path to achieving the desired vaccination targets.

\section{Methodology Limitation} 
A scholar must mention the limitations of their research as the course might shift given the variability of data. 
The main difficulties faced during the research for the development of this paper reside in the data collection, especially since Lebanon does not hold significant record-registering \& monitoring systems.
The study's aspects were limiting time management, budget allocation, lockdown restrictions, and unavailability of research regarding the topic.
Furthermore, the population sampling, the interval of time when the data was collected, and the omittance of other factors such as the vaccination center spaces, the cost of the servers, and the receptors’ ages might highly affect the results.
Lastly, the queueing theory was not previously deployed in the medical field and in this context, the sector’s parties' lack of experience constitutes a notable obstacle.

\section{Future Studies} 
Based on the results and the literature review, the subsequent studies can be developed horizontally or vertically.
A horizontal approach would be examining the here-neglected factors such as the vaccination centers’ spaces, the servers’ costs \& the receptors’ ages.
On the other hand, a vertical development would imply conducting a study concerning the queueing theory's employment in other medical fields like the emergency sections of hospitals \& diverse sectors.

\bibliographystyle{plain}
\bibliography{main}

\end{document}